\documentclass[12pt]{article}

\usepackage{amssymb}
\usepackage{amsmath}
\usepackage{array} 
\usepackage{epsfig}

\begin{document}


\begin{center}

 {\Large \bf
\vskip 7cm
\mbox{Patterns of the Exclusive Double Diffraction}
}
\vskip 1cm

\mbox{Petrov~V.A. and Ryutin~R.A.}

\mbox{{\small Institute for High Energy Physics}}

\mbox{{\small{\it 142 281} Protvino, Russia}}

 \vskip 1.75cm
{\bf
\mbox{Abstract}}
  \vskip 0.3cm

\newlength{\qqq}
\settowidth{\qqq}{In the framework of the operator product  expansion, the quark mass dependence of}
\hfill
\noindent
\begin{minipage}{\qqq}
We consider Exclusive Double Diffractive Events (EDDE) as
a powerfull tool to study the picture of the $pp$ 
interaction. Calculations of the cross-sections 
for the process $p+p\to p+M+p$ are presented in the
convenient form for further experimental 
applications. We propose measurements of t-distributions in 
the joint CMS-TOTEM experiment. It is shown that 
important information on the interaction region
could be extracted from the diffractive pattern.

\end{minipage}
\end{center}


\begin{center}
\vskip 0.5cm
{\bf
\mbox{Keywords}}
\vskip 0.3cm

\settowidth{\qqq}{In the framework of the operator product  expansion, the quark mass dependence of}
\hfill
\noindent
\begin{minipage}{\qqq}
Exclusive Double Diffractive Events -- Pomeron -- Regge-Eikonal model -- Higgs -- Radion -- Jets -- Diffractive Pattern
\end{minipage}

\end{center}

\setcounter{page}{1}
\newpage


\section{Itroduction}

 With the first LHC run coming closer the hopes for
 confirmation of various theory predictions get heated. The huge
amount of works is related to the search of fundamental particles
of the Standardt Model or its extensions (Higgs boson, Superpartners,
gravitons and so on) and to the investigations of so called "hard" QCD 
processes, which correspond to very short space-time scales. "Soft" 
diffractive proceses take in this raw its own, distinctive place.

LHC collaborations aimed at working in low and high $p_T$
regimes related to typical undulatory (diffractive) and corpuscular (point-like) 
behaviours of the corresponding cross-sections may offer a very exciting possibility to
observe an interplay of both regimes~\cite{diff1}. In theory the "hard part" can be
(hopefully) treated with perturbative methods whilst the "soft" one is definitely
nonperturbative.

 Below we give several examples of such an interplay: exclusive particle production by diffractively scattered protons, i.e. the processes $p+p\to p+M+p$, where $+$ 
 means also a rapidity gap and M represents a particle or a system of particles consisting of or strongly coupled to the two-gluon state~\cite{menu}.

 These processes are related to the dominant amplitude of exclusive 
and semiinclusive two-gluon production. Driving mechanism of the 
diffractive processes is the Pomeron. Data on the total cross-sections demands unambiguosly for the Pomeron with larger-than-one intercept, thereof the need 
to take into account the "soft" rescattering (i.e. "unitarisation"). 

 EDDE gives us unique experimental possibilities for particle searches
and investigations of diffraction itself. This is due to several advantages of the process: a) clear
signature of the process; b) possibility to use "missing mass method" that improve the mass 
resolution; c) background is strongly suppressed; d) spin-parity analysis of the central system can be 
done; e) interesting measurements concerning the interplay between "soft" and "hard" scales are possible. All these properties can be realized
in common CMS/TOTEM detector measurements at LHC~\cite{TDR:TOTEM}.

\section{Exclusive double diffraction}

The exclusive double diffractive process is related to the
dominant amplitude of the exclusive two-gluon production. Driving
mechanism of this processes is the Pomeron.

To calculate an amplitude of the EDDE, we use an
approach which was considered in detail in Ref.~\cite{menu}.
In the framework of this approach, the amplitude can be sketched
as shown in Fig.~1. After the tensor contraction of
the amplitudes $T_{1,2}$ with the gluon-gluon fusion vertex, the
full ``bare'' amplitude $T_M$ depicted in Fig.~1
looks like
\begin{equation}\label{Tpp_pXp}
T_M = \frac{2}{\pi} \, c_{gp}^2 \, e^{b(t_1+t_2)}
\left(-\frac{s}{M^2}\right)^{\alpha_P(0)} F_{gg \to M} \, I_s \;.
\end{equation}
Here
\begin{eqnarray}\label{slope}
b &=& \alpha^{\prime}_P(0) \ln \left( \frac{\sqrt{s}}{M} \right)
+ b_0 \; ,
\\
b_0 &=& \frac{1}{4} \, (\frac{r^2_{pp}}{2} + r^2_{gp}) \; ,
\end{eqnarray}
with the parameters of the ``hard'' Pomeron trajectory, that
appears to be the most relevant in our case, presented in
Table~\ref{tab:hpomeron}. The last factor in the r.h.s. of
~(\ref{Tpp_pXp}) is
\begin{eqnarray} \label{Isudakov}
I_s &=& \int\limits_{0}^{\mu^2}\frac{dl^2}{l^4} \, F_s(l^2,\mu^2)
\left( \frac{l^2}{s_0 + l^2/2} \right)^{2\alpha_P(0)}\; ,
\end{eqnarray}
where $l^2=-q^2\simeq{\mbox{\bf q}}^2$, $\mu=M/2$, and $s_0$ is a scale
parameter of the model which is also used in the global fitting of
the data on $pp$ ($p\bar{p}$) scattering for on-shell
amplitudes~\cite{diff1}. The fit gives $s_0 \simeq 1$~GeV$^2$.
If we take into account the emission of virtual "soft" gluons,
while prohibiting the real ones, that could fill rapidity gaps, it
results in a Sudakov-like suppression~\cite{KMR3:sudakov}:
\begin{eqnarray}\label{sudakov}
F_s(l^2,\mu^2) &=& \exp\left[ -
\int\limits_{l^2}^{{\mu}^2} \frac{d p_T^2}{p_T^2} \,
\frac{\alpha_s({p_T}^2)}{2\pi} \int\limits_{\Delta}^{1-\Delta} z P_{gg}(z) d z
+\int\limits_0^1\sum_q P_{qg}(z)dz
\right] \; ,\\
P_{gg}(z)&=&6\frac{(1-z(1-z))^2}{z(1-z)}\; ,\\
\Delta&=&\frac{p_T}{p_T+\mu}\; .
\end{eqnarray}

\begin{figure}[t]
\label{fig:EDDE}
\hskip  5cm \vbox to 4cm 
{\hbox to 6cm{\epsfxsize=6cm\epsfysize=4cm\epsffile{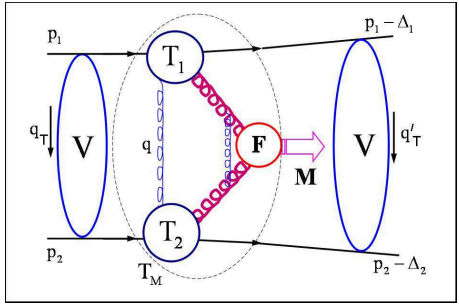}}}
\caption{Model for EDDE}
\end{figure}

The off-shell gluon-proton amplitudes $T_{1,2}$ are obtained in
the extended unitary approach~\cite{Petrov:95}. The ``hard'' part
of the EDDE amplitude, $F_{gg\to M}$, is the usual gluon-gluon
fusion amplitude calculated perturbatively in the SM or in its
extensions.

\begin{figure}[bt]
\label{fig:CDFdata}
\hskip  5cm \vbox to 4cm 
{\hbox to 5cm{\epsfxsize=5cm\epsfysize=4cm\epsffile{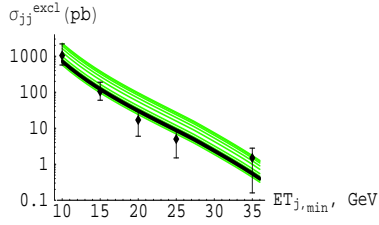}}}
\caption{The latest data from CDF and predictions for EDDE.}
\end{figure}

\begin{table}[h!]
\begin{center}
\caption{Phenomenological parameters of the ``hard'' Pomeron
trajectory obtained from the fitting of the HERA and Tevatron data
(see~\cite{menu}, \cite{EVMP:HERA}), and data on $pp$
($p\bar{p}$) scattering~\cite{diff1}.The value of the $c_{gp}$ is corrected
in accordance with the latest data from CDF~\cite{CDF2006}, which
is depicted in the Fig.~2 with the range of possible curves.}
\bigskip \bigskip
  \begin{tabular}{||c|c|c|c|c||}
  \hline
  $\alpha_P(0)$ & $\alpha_P^{\prime}(0)$, GeV$^{-2}$ &  $r^2_{pp}$, GeV$^{-2}$
  & $r^2_{gp}$, GeV$^{-2}$ & $c_{gp}$
  \\ \hline
  1.203 &  0.094 &  2.477 &  2.54 & 3.2$\pm$0.5
  \\ \hline
  \end{tabular}
\label{tab:hpomeron}
\end{center}
\end{table}

The data on total cross-sections demand unambiguously the Pomeron
with larger-than-one intercept, thereof the need in unitarization.
The amplitude with unitary corrections, $T^{unit}_M$, are depicted
in Fig.~1. It is given by the following analytical
expressions:
\begin{eqnarray}\label{ucorr}
T^{unitar}_M(p_1, p_2, \Delta_1, \Delta_2) &=& \frac{1}{16\,s
s^{\prime}} \int \frac{d^2\mbox{\bf q}_T}{(2\pi)^2} \,
\frac{d^2\mbox{\bf q}^{\prime}_T}{(2\pi)^2} \; V(s, \mbox{\bf
q}_T) \;
\nonumber \\
&\times& T_M( p_1-q_T, p_2+q_T,\Delta_{1T}, \Delta_{2T}) \,
V(s^{\prime}, \mbox{\bf q}^{\prime}_T) \;,
\\
V(s, \mbox{\bf q}_T) &=& 4s \, (2\pi)^2 \, \delta^2(\mbox{\bf
q}_T) + 4s \!\! \int d^2\mbox{\bf b} \, e^{i\mbox{\bf q}_T
\mbox{\bf b}} \left[e^{i\delta_{pp\to pp}}-1\right]\;,
\end{eqnarray}
where $\Delta_{1T} = \Delta_{1} -q_T - q^{\prime}_T$, $\Delta_{2T}
= \Delta_{2} + q_T + q^{\prime}_T$, and the eikonal function
$\delta_{pp\to pp}$ can be found in Ref.~\cite{diff1}. Left
and right parts of the diagram in Fig.~1b denoted by
$V$ represent ``soft'' re-scattering effects in initial and final
states, i.e. multi-Pomeron exchanges. As was shown in
\cite{EDDE:glueballs}, these ``outer'' unitary corrections
strongly reduce the value of the corresponding cross-section and
change an azimuthal angle dependence.

 In the equation~(\ref{Tpp_pXp}) we present only the Born terms from
 amplitudes $T_{1,2}$. It is sufficient for $|t_{1,2}|<1$~GeV due to fast
 decrease of the differential cross-section in $t_{1,2}$, and the 
 contribution of these corrections to the total cross-section are less than
 several percents. But when
we consider the diffractive pattern in the region 
of $1<|t_{1,2}|<5$~GeV, we have to take
into account rescattering corrections inside the amplitudes $T_{1,2}$. In this case
$I_s$ in the 
equation~(\ref{Tpp_pXp}) changes to the following expression:
\begin{eqnarray}\label{T12cor}
I^{corr}_s &=& \int\limits_{0}^{\mu^2}\frac{dl^2}{l^4} \, F_s(l^2)
\left( \frac{l^2}{s_0 + l^2/2} \right)^{\alpha_P(t_1)+\alpha_P(t_2)}
(1+h(v,t_1))
(1+h(v,t_2))\; ,\\
\label{T12cor_h}
h(v,t)&=& \sum_{n=2}^{\infty} \frac{(-1)^{n-1}}{n!\cdot n}
\left(  \frac{c_{gp}}{8\pi\; b_1(v)}  
\exp\left[-\frac{i\pi(\alpha_P(0)-1)}{2} \right] v^{\alpha_P(0)-1} 
\right)^{n-1}\cdot\nonumber\\
&\cdot&\exp\left[ \frac{b_1(n-1)}{n}|t| \right]\; ,\\
v&=& \frac{\sqrt{s}}{M}\frac{l^2}{s_0 + l^2/2}\; ,
\end{eqnarray}
and $b$ to
\begin{equation}
\label{T12cor_b1}
b_1= \alpha^{\prime}_P(0) \ln v +b_0\; .
\end{equation}

To calcuate differential and total cross-sections for exclusive processes
we can use the formula
\begin{eqnarray}
\label{eq:diffcs}
M^2\frac{d\sigma^{EDDE}}{dM^2\; dy\; d\Phi_{gg\to M}}|_{y=0}&=&\hat{L}^{EDDE}
\frac{d\hat{\sigma}^{J_z=0}}{d\Phi_{gg\to M}}\; ,\\
\label{lumEDDE}
\hat{L}^{EDDE}&=& \frac{c_{gp}^4}{2^5\pi^6}
\left( \frac{s}{M^2} \right)^{2(\alpha_P(0)-1)} \frac{1}{4b^2} I_s
S^2\; ,\\
\label{softsurv}
S^2 &=& \frac{\int d^2\vec{\Delta}_1d^2\vec{\Delta}_2 |T^{unitar}_M|^2}{\int d^2\vec{\Delta}_1d^2\vec{\Delta}_2 |T_M|^2}\; ,
\end{eqnarray}
where $d\hat{\sigma}^{J_z=0}/d\Phi_{gg\to M}$ is the "hard" 
exclusive singlet gluon-gluon fusion cross-section and $S^2$ is the so called "soft" 
survival probability. In this work the quantity $\hat{L}$ is called $g^{IP}g^{IP}$ luminocity.

\begin{figure}[t]
\label{fig:softsurv}
\hskip  5cm \vbox to 4cm 
{\hbox to 6cm{\epsfxsize=6cm\epsfysize=4cm\epsffile{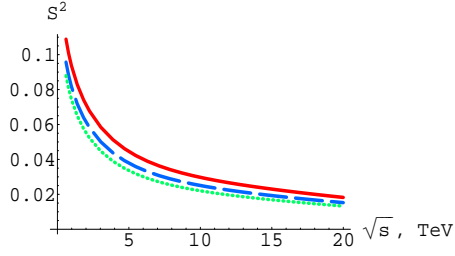}}}
\caption{"Soft" survival probability $S^2$ as a function of $\sqrt{s}$ for masses
of the central system $10$~GeV (solid curve), $50$~GeV (dashed one) and $200$~GeV (dotted one).}
\end{figure}

The factor $S^2$ is depicted in the Fig.~3 for the systems M with quantum numbers 
$0^{++}$ (SM Higgs boson, Radion, jet-jet). For other cases it is of the same order and
can be calculated using the Monte-Carlo event generator EDDE~\cite{ManualEDDE}.

\section{Results}

\begin{figure}[t]
\label{fig:gammas}
\vbox to 10cm 
{\hbox to 14cm{\epsfxsize=14cm\epsfysize=10cm\epsffile{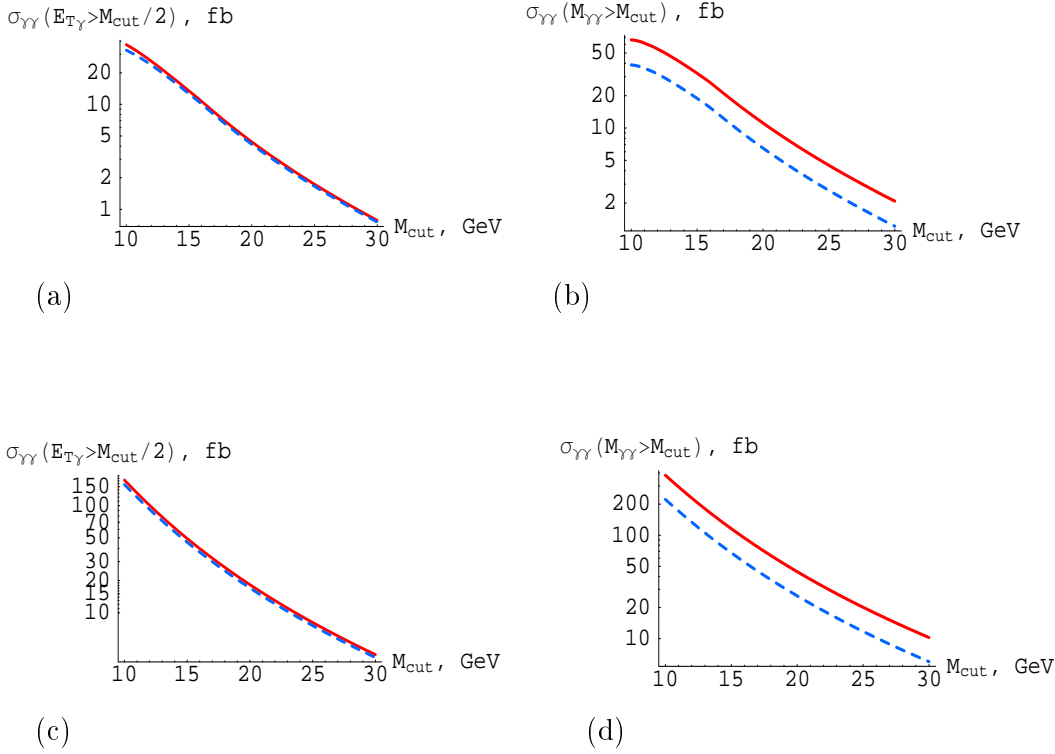}}}
\caption{Cross-sections for the process $pp\to p+\gamma\gamma+p$ for different kinematical cuts. Solid and dashed curves correspond to the pseudorapidity cuts $|\eta_{\gamma}|<2$ and $|\eta_{\gamma}|<1$. a) $\sqrt{s}=1.8$~TeV, CDF cuts for $\xi_{1,2}$~\cite{CDF2006}, and cut on the $E_{T\gamma}$; b) $\sqrt{s}=1.8$~TeV, CDF cuts for $\xi_{1,2}$~\cite{CDF2006}, and cut on the $M_{\gamma\gamma}$; c) $\sqrt{s}=14$~TeV, symmetric cuts $0.0003<\xi_{1,2}<0.1$, and cut on the $E_{T\gamma}$; d) $\sqrt{s}=14$~TeV, symmetric cuts $0.0003<\xi_{1,2}<0.1$, and cut on the $M_{\gamma\gamma}$.}
\end{figure}

First of all we would like to discuss some features of the process $pp\to p+\gamma\gamma+p$, since this process is the standard one to obtain the model parameters. Cross-sections for this process are presented in the Fig.~4. It is important
to note that the cut $E_{T\gamma}>E_{cut}=M_{cut}/2$ is used in the major part
of experimental works, that is why we have to use the same one in our 
calculations.  But in some theoretical works~\cite{khoze2gamma} $E_{T}>E_{cut}$ 
means another cut $M_{\gamma\gamma}>2E_{cut}$, which leads to the result, similar
to the one presented in the Fig.~4b. In this figure cross-section for
$|\eta_{\gamma}|<2$ is about two times higher than for $|\eta_{\gamma}|<1$. Such 
difference is only possible in the kinematics, when 
$M_{\gamma\gamma}>2E_{cut}$. It follows from rather simple calculations. 
Total cross-section for the process $gg\to\gamma\gamma$ can be
represented as~\cite{TTWu}
\begin{equation}
\label{cs-partonic}
\hat{\sigma}^{J_z=0}_{gg\to\gamma\gamma}(M_{\gamma\gamma},\eta_{max})=C_{\gamma\gamma}F(\eta_{max})\frac{\alpha_s(M_{\gamma\gamma}/2)^2}{M_{\gamma\gamma}^2}\;,\;
\end{equation}
where $\eta_{max}$ is the pseudorapidity cut in the central mass frame of the 
diphoton system, $C_{\gamma\gamma}$ is the constant, 
\begin{equation}
F(\eta_{max})=\int_{-\eta_{max}}^{\eta_{max}} \frac{d\eta}{{\rm ch}^2\eta} \left[
1+\left(
1-2\eta\;{\rm th}\;\eta+\frac{1}{4}\left( \pi^2+4\eta^2\right)\left( 1+{\rm th}^2\eta\right)
\right)^2
\right]
\end{equation}
is depicted in the Fig.~5a. 
\begin{figure}[t]
\label{fig:Ffunction}
\hskip 1cm \vbox to 5cm 
{\hbox to 12cm{\epsfxsize=12cm\epsfysize=5cm\epsffile{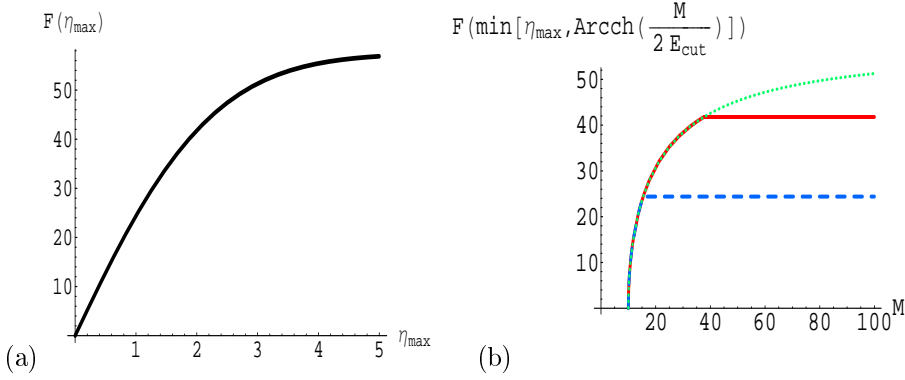}}}
\caption{a) Function $F(\eta_{max})$; b) $\eta_{max}=2$ (solid curve), $\eta_{max}=1$ (dashed one), $F({\rm Arcch}\frac{M}{2E_{cut}})$ (dotted one), $E_{cut}=5$~GeV.}
\end{figure}
And for the process $pp\to p+\gamma\gamma+p$ 
from~(\ref{eq:diffcs}) we have
\begin{equation}
\label{cs-hadronic}
\sigma_{pp\to p+\gamma\gamma+p}(E_{cut},\eta_{max})\simeq\int_{2E_{cut}}^{\sqrt{\xi_{1max}\xi_{2max}s}}
\frac{dM^2}{M^2}\hat{L}^{EDDE}(M)\hat{\sigma}^{J_z=0}_{gg\to\gamma\gamma}(M,\eta_{max})\Delta y\;.
\end{equation}

We are interested in the ratio of total cross-sections for different $\eta_{max}$. Let us
consider first the kinematics with cuts
\begin{equation}
\label{eq:Mcut}
M_{\gamma\gamma}>2E_{cut}\;,\; |\eta_{\gamma}|<\eta_{max}\;.
\end{equation}
In this case 
\begin{equation}
\label{eq:ratioM}
\frac{\sigma_{pp\to p+\gamma\gamma+p}(M>2E_{cut},|\eta|<2)}{\sigma_{pp\to p+\gamma\gamma+p}(M>2E_{cut},|\eta|<1)}\simeq\frac{F(2)}{F(1)}\simeq 1.7\;.
\end{equation}
Since in the central mass frame of the diphoton system we have 
$M_{\gamma\gamma}=2E_{T\gamma}{\rm ch}\;\eta_{\gamma}$, in 
the kinematics with
\begin{equation}
\label{eq:ETcut}
E_{T\gamma}=\frac{M_{\gamma\gamma}}{2\;{\rm ch}\;\eta_{\gamma}}>E_{cut}\;,\; |\eta_{\gamma}|<\eta_{max}
\end{equation}
we have additional cut 
\begin{equation}
\label{eq:addcuteta}
|\eta_{\gamma}|<{\rm Arcch}\;\frac{M_{\gamma\gamma}}{2E_{cut}}\;,
\end{equation}
and we should use $F\left( \min\left[ \eta_{max}, {\rm Arcch}\;\frac{M_{\gamma\gamma}}{2E_{cut}}\right]\right)$ instead of $F(\eta_{max})$ in~(\ref{cs-partonic}). This new function is shown in the Fig.~5b.  The main contribution to the integral
comes from small masses due to fast decrease in $M$, but in this region we have the same cut $|\eta|<{\rm Arcch}\frac{M}{2E_{cut}}$ for different $\eta_{max}$. And it is
easy to get the following result
\begin{equation}
\label{eq:ratioET}
\frac{\sigma_{pp\to p+\gamma\gamma+p}(E_T>E_{cut},|\eta|<2)}{\sigma_{pp\to p+\gamma\gamma+p}(E_T>E_{cut},|\eta|<1)}\simeq 1.1\; \mbox{for}\; E_{cut}=5\;\mbox{GeV}\;.
\end{equation}
Even if we take $\alpha_s=const$ and $\hat{L}^{EDDE}=const$, we will get the ratio $1.3$, and not $\sim 2$ as in~\cite{khoze2gamma}. This simple example shows that we
should be carefull with the kinematics during our calculations, since this could lead
to different predictions. Now we can compare our result with the latest data on
the exclusive $\gamma\gamma$ production
from CDF~\cite{AlbrowCMSweek}
\begin{equation}
\label{dataAlbrow}
\sigma_{pp\to p+\gamma\gamma+p}(E_{T\gamma}>5\;\mbox{GeV},|\eta_{\gamma}|<2)=
0.14{+0.14\atop -0.04}\;\mbox{(stat.)}\;\pm 0.03 \;\mbox{(syst.)}\; \mbox{pb}\;.
\end{equation}
This prediction is higher than our calculations based on the di-jet production~\cite{CDF2006}. If we beleive in this result,
the model parameter $c_{gp}$ is at least 20\% higher, than our 
estimations. The possible reason is that in $\gamma\gamma$ production 
we use the region of small masses
for the normalization of our 
parameter (and higher masses for the di-jet production), 
but the uncertainty in the scale 
dependence of the cross-section is rather large (factor $\sim 2$). This
result from CDF may serve a good signal for the future exclusive Higgs
boson production, since it makes the cross-section higher by about two times.

 Now we can estimate the backgrounds for the exclusive Higgs production at LHC~\cite{menu}. Rates
for these processes at the integrated luminocity $100$~fb$^{-1}$ are summarized in the table~\ref{tab:Higgs} (parameter $c_{gp}=3.7$, i.e. we use more optimistic
values based on the data on the di-jet production~\cite{CDF2006}). From this table we can obtain signal to background ratio $\sim 1$. More exact estimations will be made in the nearest future after full Monte-carlo simulations.

\begin{table}[hbt]
\begin{center}
\caption{Rates for the exclusive Higgs production and different backgrounds
at the integrated luminocity $100$~fb$^{-1}$ and $\Delta M_{missing}=4$~GeV. The probability to misidentify gluon jets with b-jets is taken to be 1\%.}
\bigskip \bigskip
  \begin{tabular}{|c|c|}
  \hline
  process & N events
  \\ \hline
  $\sigma^{EDDE}\left(H\to b\bar{b}\right)$ &  27
  \\ \hline
$\sigma^{SI}\left(H\to b\bar{b}\right)$  &  2
  \\ \hline
$\sigma^{EDDE}\left(b\bar{b}\right)$  &  12
  \\ \hline
  $\sigma^{EDDE}\left(b\bar{b}g\right)$ &  1
  \\ \hline  
  $\sigma^{EDDE}\left(gg\right)\cdot 10^{-4}$ &  14
  \\ \hline
  $\sigma^{EDDE}\left(ggg\right)\cdot 10^{-4}$ &  2
  \\ \hline  
  \end{tabular}
\label{tab:Higgs}
\end{center}
\end{table}

\section{Diffractive patterns}

In the general agenda of the LHC experiment diffraction often looks 
as an "auxiliary tool" for other processes
such as Higgs boson and exotics searches, background supression and so 
on. Nevertheless, diffractive measurements have their own classical tasks
directly related to the angular (or~$t$) distributions.

 Diffractive pattern is usually characterized by the peak at
small values of $t$, and complicated structure with 
dips or breaks and bumps for larger $t$~\cite{diff1}.  This picture 
reflects the ondulatory properties of quantum processes as contrasted 
to more habitual particle-like behaviour, and allows us to get an information 
about the size and shape of the strong interaction region at large 
distances (i.e. directly related to confinement of the QCD colour fields).

\begin{itemize}
\item From the diffractive pattern we extract model independent parameters
of the interaction region such as the $t$-slope which is $R^2/2$, with $R$ the transverse radius of the interaction region.
\item We can also estimate the longitudinal size of the interaction region~\cite{diff2}:
\begin{eqnarray}
\label{interreg}
&&\Delta x_L>\frac{\sqrt{s}}{2\sqrt{<t^2>-<t>^2}}
\end{eqnarray}
The longitudinal interaction range is somehow "hidden" in the amplitude
but it is this range that is responsible for the "absorption strength". A
rough analogue is the known expression for the radiation absorption
in media which critically depends on the thickness of the absorber.
\item The very presence of dips is the signal of the quantum 
interference of hadronic waves.
\item The depth of dips is determined by the real part of the scattering
amplitude
\end{itemize}

According to the data from $Sp\bar{p}S$ and Tevatron the transverse 
radius of the interaction region is of order 
of $1.2 \;{\rm fm}\simeq 1.5<r_{\rm em}>$. The longitudinal size can 
be estimated from the second inequality~(\ref{interreg}) and is of order 
of 2800~fm.

Diffractive pattern moves due to changes in kinematical
parameters like the energy of the interaction or an additional hard scale.
This motion reflects the dynamics of the process.
The increase of the $t$-slope with energy reflects the growth of the 
interaction radius. At fixed collision energy the diffractive pattern is 
fixed as well. 

However if we have in our disposal an additional hard 
scale we can operate the diffractive pattern adjusting this hard scale 
at our will and making, e.g., the interaction region larger or smaller. 

Hard scale is related to small distances and,
from the simple optical point of view, the pattern should move towards large
values of -$t$ with the increase of the hard scale. HERA provides an 
excellent opportunity to observe the influence of a hard scale ($Q^2$) 
on the diffractive pattern: the slope decreases with $Q^2$ in exclusive vector 
meson or photon production or for mesons 
containing heavy quarks ($J/\Psi$) as contrasted to those composed of 
the light quarks (see Fig.~6)~\cite{HERAslope}.

\begin{figure}[hbt]
\label{figb}
\hskip  6cm \vbox to 6cm 
{\hbox to 4cm{\epsfxsize=4cm\epsfysize=6cm\epsffile{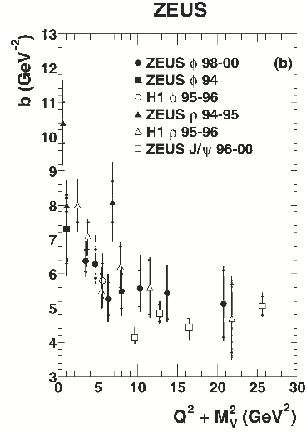}}}
\caption{The slope b, as a function of $Q^2+M_V^2$, compared to 
other ZEUS and H1 results.}
\end{figure}

\begin{figure}[hbt]
\label{petrovfig}
\hskip  4cm \vbox to 4cm 
{\hbox to 6cm{\epsfxsize=6cm\epsfysize=4cm\epsffile{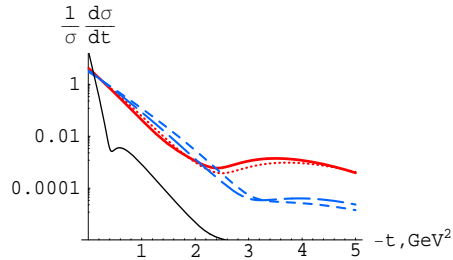}}}
\caption{Normalised cross section for exclusive 
dijet production as a function of $t$ for $M_X=30$~GeV 
(the solid and long-dashed curves correspond to the LHC and TEVATRON 
energies, respectively) and $M_X=200$~GeV (the dotted and short-dashed 
curves correspond to the LHC and TEVATRON 
energies, respectively).The left curve corresponds to the elastic scattering 
at the LHC.}
\end{figure}

The decrease of the slope
with $Q^2$ in electroproduction was predicted qualitatively in Ref.~\cite{diff4}:
J.D. Bjorken argued that the decrease of the slope would be bounded
from below by the size of nucleon~\cite{diff5}. The latter feature seems 
to be violated in the HERA data~\cite{diff6}. 
We have to mention that the presence of a high-mass particle in the final 
state does not always lead to the phenomena described above. For 
example, hadronic resonances with large 
masses have large size due to intrinsic motion of constituents, and can 
not be considered as a hard probe. In this case we have inverse dynamics 
of the pattern~\cite{lowres}. This certainly is not the 
case for the processes considered below as they are related exclusively 
to short-distance probes, i.e. "high mass" means 
always "high ${\rm E}_{\rm T}$".

The diffractive pattern for the process $p+p\to p+jj+p$ as predicted on 
the basis of ref.~\cite{menu} is displayed on the Fig.~7 
where $\frac{1}{\sigma}\frac{d\sigma}{dt}$ means the exclusive 
differential cross section with all final variables integrated except 
one of the proton transverse momenta ($-t$) and the value of the 
central mass (${\rm M}\simeq 2{\rm E}_{\rm T}$). With two exclusive 
high-${\rm E}_{\rm T}$ jets the expected dips will reflect the elastic 
scattering of the protons off the hard gluon. Their 
positions are shifted to the right in comparison with the proton-proton elastic 
scattering, as depicted on the Fig.~7. Such a shift is a clear signal of the 
short-distance scale due to jets. 

 Measurements of t-distributions and their dynamics 
in the exclusive central diffraction
could be used for the proposed investigations. To obtain the detailed 
diffractive pattern
with dips  for 1 GeV$^2<-t<$ 5 GeV$^2$ we need at least $10^4$ 
events for fixed (or falling within the small enough range of values) 
masses of the central
system and $t$-resolution 
less than 10\% in this region. At high
luminosities the use of the missing mass method is limited below
by central masses above 30~GeV because of the acceptance limitations
and the absence of resonances with high rates in this region. That is why 
the only way is to use exclusive or Semi-Inclusive (exclusive+"soft"
radiation in the central rapidity region) dijet production. The best
case is the measurements at the nominal luminosity at $\beta^*=0.5$. Results
are summarized in the table~\ref{diffpattrates}. 
\begin{table}[h!]
\caption{\label{diffpattrates} Rates for exclusive and
semi-inclusive ($|\eta_{soft}|<5$) double diffractive dijet production
for luminosity $10^{33}$~cm$^{-2}$~s$^{-1}$ for different intervals of the 
invariant mass of the central system, $M_X$.}
\begin{center}
\begin{tabular}{|c|c|c|c|}
\hline
   M$_1<$ M$_X<$ M$_2$(GeV)  & $t$-slope (GeV$^{-2}$) & N$_{\rm ex}$  & N$_{\rm semi-incl.}$\\
\hline 
      29$<$ M$_X<$31        &         4.6    &  $2\cdot 10^4/$day  &  $6\cdot 10^4/$day \\
\hline 
      98$<$ M$_X<$102      &         4.3    &   $9\cdot 10^3/$month  &  $4.5\cdot 10^4/$month \\
\hline 
    196$<$ M$_X<$204       &        4.1     & $5.5\cdot 10^3/$year &   $4\cdot 10^4/$year \\
\hline 
\end{tabular}
\end{center}
\end{table}


\section*{Aknowledgements}

This work is supported by grants RFBR-06-02-16031 and INTAS-05-112-5481.

\end{document}